\documentclass{aastex6}

\usepackage{soul}
\usepackage{rotating}

\usepackage{color}

\usepackage{multirow}
\usepackage{amsmath}	
\usepackage{amssymb}	
\usepackage{gensymb}
\usepackage{subfigure}

\begin{document}

\title{
A Sequoia in the Garden: FSR\,1758 -  Dwarf Galaxy or Giant Globular Cluster? 
\footnote{Based  on  observations  taken  within  the  ESO  programmes  179.B-2002, and 198.B-2004.}}

\author{
\and Rodolfo H. Barb\'a\altaffilmark{1},
\and Dante Minniti\altaffilmark{2,3,4},
\and Douglas Geisler\altaffilmark{5,1,7},
\and Javier Alonso-Garc\'ia\altaffilmark{6,3},
\and Maren Hempel\altaffilmark{2},
\and Antonela Monachesi\altaffilmark{7,1},
\and Julia I. Arias\altaffilmark{1},
\and Facundo A. G\'omez\altaffilmark{7,1}
}

\altaffiltext{1}{Departamento de F\'isica y Astronom\'ia, Universidad de La Serena, Avenida Juan Cisternas 1200, La Serena, Chile.}
\altaffiltext{2}{Depto.  de   Ciencias  F\'isicas,  Facultad   de  Ciencias  Exactas, Universidad  Andres Bello, Fernandez  Concha 700, Las  Condes, Santiago, Chile.}
\altaffiltext{3}{Millennium  Institute  of  Astrophysics,  Av.  Vicuna  Mackenna 4860, 782-0436, Santiago, Chile.}
\altaffiltext{4}{Vatican  Observatory,  V00120   Vatican  City  State,  Italy.}
\altaffiltext{5}{Departamento de Astronomia, Casilla 160-C, Universidad de Concepcion, Chile.}
\altaffiltext{6}{Centro de Astronom\'{i}a (CITEVA), Universidad de Antofagasta, Av. Angamos 601, Antofagasta, Chile.}
\altaffiltext{7}{Instituto de Investigaci\'on Multidisciplinar en Ciencia y Tecnolog\'ia, Universidad de La Serena, Ra\'ul Bitr\'an 1305, La Serena, Chile.}

\email{rbarba@userena.cl}




\begin{abstract}
We present the physical characterization of FSR\,1758, a new large,
massive object very recently discovered in the Galactic Bulge.  
The combination of optical data from the 2nd Gaia Data Release (GDR2) and the DECam Plane Survey (DECaPS), and near-IR data from the VISTA Variables in the V\'{\i}a L\'actea Extended Survey (VVVX) led to a clean sample of likely members.
Based on this integrated dataset, position, distance, reddening, 
size, metallicity, absolute magnitude, and proper motion of this object are measured. 
We estimate the following parameters: 
$\alpha=17:31:12$, $\delta=-39:48:30$ (J2000),
$D=11.5 \pm 1.0$ kpc, 
$E(J-Ks)=0.20 \pm 0.03$ mag, 
{\bf $R_c=10$ pc,
$R_t=150$ pc},  
$[Fe/H]=-1.5 \pm 0.3$ dex, 
$M_i < -8.6 \pm 1.0$, 
$\mu_{\alpha} = -2.85$ mas yr$^{-1}$, and  
$\mu_{\delta} = 2.55$ mas yr$^{-1}$.
The nature of this object is discussed. 
If FRS\,1758 is a genuine globular cluster, it is one of the largest in the Milky Way,  with a size comparable or even
larger than that of $\omega$ Cen, being also an extreme outlier in the size vs. Galactocentric distance diagram. 
The presence of a concentration of long-period RR Lyrae variable stars and blue horizontal branch stars suggests that it is a typical metal-poor globular cluster of  Oosterhoff type II.
Further exploration of a larger surrounding field reveals common
proper motion stars, suggesting either tidal debris or that FRS\,1758 is actually the
central part of a larger extended structure such as a new dwarf galaxy, tentatively named as Scorpius. 
In either case, this object is remarkable, and its discovery graphically illustrates the possibility to find other large objects hidden in the Galactic Bulge using future surveys. 
\end{abstract}


\section{Introduction} 
\label{sec:intro}

More than two dozen of new low luminosity globular cluster (GC)
candidates have been discovered in the past year in the direction to the Galactic Bulge  (Minniti et al. 2017a,b, Ryu \& Lee 2018, Bica et al. 2018, Camargo 2018,  Palma et al. 2018 in preparation). These objects are very difficult to detect, due to the heavy
extinction and high field stellar density lying well inside the Bulge, and, if proved to be genuine clusters, most are expected to be of low mass.  

Very recently, Cantat-Gaudin et al. (2018), on the basis of the Gaia optical  color-magnitude diagram (CMD), proposed another object, [FSR2007] 1758 (in short FSR\,1758), to be a new GC located towards the Bulge. 
This serendipitous discovery was made while studying 1229 open
clusters and noticing  the striking proper motion (PM) separation from the field stellar population.  
They estimated a Galactocentric distance $R_{G}=1600$ pc, and a height below the plane $z=-470$ pc, which places it inside
the Bulge itself.
This object was listed as a diffuse open cluster in the catalogs of
Froebrich et al. (2007) and Kharchenko et al. (2013), although we note that the position and physical properties  in these initial studies are very preliminary. 

In this paper, we use combined data from 2nd Gaia Data Release (GDR2, Gaia Collaboration 2018), the DECaPS Survey (Schlafly et al. 2018), and the VVVX Survey \citep{minniti18a} to investigate the physical properties of this impressive object in much greater detail.
We also use OGLE RR Lyrae stars to provide an external
distance determination.
Section~\ref{sec:sec2}, \ref{sec:sec3} and \ref{sec:sec4} describe the discovery, observations used, and derived color-magnitude diagrams, respectively.
Then, based on our findings, we discuss its physical nature in Section~\ref{sec:sec5}  and summarize some conclusions in Section~\ref{sec:sec6}.

\section{Independent discovery of FSR\,1758}
\label{sec:sec2}

\begin{figure}[ht]
\centering
\includegraphics[height = 100 mm]{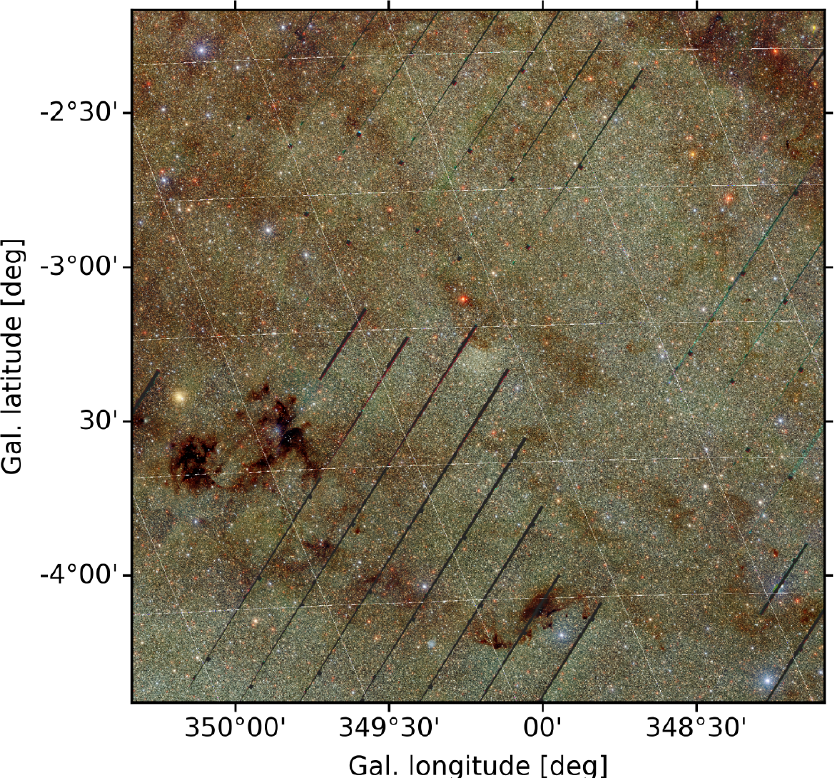}
\includegraphics[height = 100 mm]{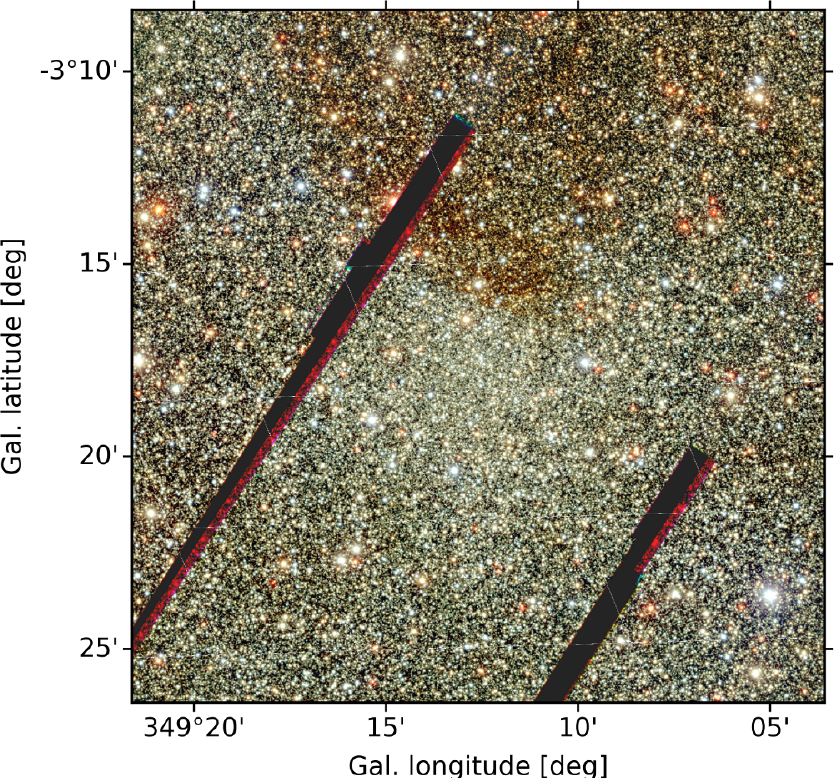}
\caption{ 
Optical finding charts for the new GC from the DECaPS survey (Schlafly
et al. 2018), showing a field of about $2.25\deg \times 2.25\deg$
(left), and a zoomed-in region of about $18' \times 18'$  (right),
both centered on the new globular cluster and in Galactic
coordinates. 
}
\label{fig:DECaPS-images}
\end{figure}

One of us (R.B.) made the independent discovery of FSR\,1758 
serendipitously by visual inspection of the images from the DECaPS
Survey of Schlafly et al. (2018).
It stands out as a  bright, diffuse glow in an otherwise patchy and
generally high extinction zone, next to an extended dark cloud (Figure~\ref{fig:DECaPS-images}). 
A zoom reveals that the diffuse object is really a plethora of stars.
A quick inspection of the GDR2 PMs in the region revealed that the
cluster motion is very different from the field stars, as also noted by Cantat-Gaudin et al. (2018).

FSR\,1758 appears to be quite large, perhaps rivaling in size  with the largest Galactic GCs $\omega$ Cen and NGC\,2419 (e.g. Harris 1996, Ripepi et al. 2007). 
The visible part seen in the DECaPS images (Figure~\ref{fig:DECaPS-images}) is probably just "the tip of the iceberg", being much of its population possibly hidden by field contamination and differential reddening. 
Part of the area has surprisingly low reddening, and it is this region that we use to obtain the best cluster CMDs in order to determine its parameters.  
Nevertheless, approximately one third of the object is heavily obscured by one or more interstellar clouds towards the plane.
Thus, evidence suggests that the cluster may be more extended and/or have tidal tails, as discussed below.

\section{GDR2, DeCAPS and VVVX Data}
\label{sec:sec3}

Because of the large non-uniform reddening in the bulge fields studied here  \citep[e.g.][]{schlafly11, gonzalez12, minniti18b, alonso15, alonso18}, we also use the  combination of optical GDR2 with near-IR VVVX data to help determine the parameters. 

The GDR2 \citep{gaia18} significantly improves the photometric and astrometric measurements of the first release, and provides additional information on astrophysical parameters, variability and median radial velocities for some sources. 
The GDR2 data have been processed by the {\it Gaia Data Processing and Analysis Consortium (DPAC)}, and  contains  $G$-band magnitude
for over $1.6\times 10^9$ sources with $G < 21$ mag, and broad band colours $G_{BP}$ covering $(330-680~nm)$ and $G_{RP}$ covering $(630-1050~nm)$, for $1.4\times 10^9$ sources.
PM components in equatorial coordinates are available for $1.3 \times 10^9$ sources, with an accuracy of $0.06$ mas yr$^{-1}$, $0.2$ mas yr$^{-1}$, and $1.2$ mas yr$^{-1}$, for sources with $G < 15$ mag, $G\sim 17$ mag, and $G\sim 20$ mag, respectively \citep{gaia18}.

DECaPS (Schlafly et al. 2018) is an optical multi-band survey of the Southern Galactic Plane performed with the DECAM camera attached to the Victor Blanco 4 m telescope at Cerro Tololo Inter-American Observatory (Chile).

The VVVX survey (Minniti 2018) maps the Galactic Bulge and southern
disk in the near-IR with the VIRCAM (VISTA InfraRed CAMera) at the 4.1 m wide-field Visible and Infrared Survey Telescope for Astronomy \citep[VISTA,][]{emerson10} at ESO Paranal Observatory (Chile).  
In the Galactic Bulge, the VVVX Survey covers about 600~${\rm deg}^2$, using the $J$ (1.25~$\mu$m), $H$ (1.64~$\mu$m), and $Ks$ (2.14~$\mu$m) near-IR passbands. 
The VVVX Survey data reduction and the archival merging were carried out at the Cambridge Astronomical Survey Unit \citep[CASU,][]{irwin04} and VISTA Science Archive (VSA) at the Wide-Field Astronomy Unit (WFAU), within the VISTA Data Flow System \citep{cross12}.  
In order to deal with the high crowding in this VVVX,  we follow
\citet{alonso18}, extracting the PSF photometry and obtaining a highly complete near-IR catalog. 

\section{Color-Magnitude Diagrams}
\label{sec:sec4}

\begin{figure}[ht]
\centering
\includegraphics[height=150mm,angle=-90]{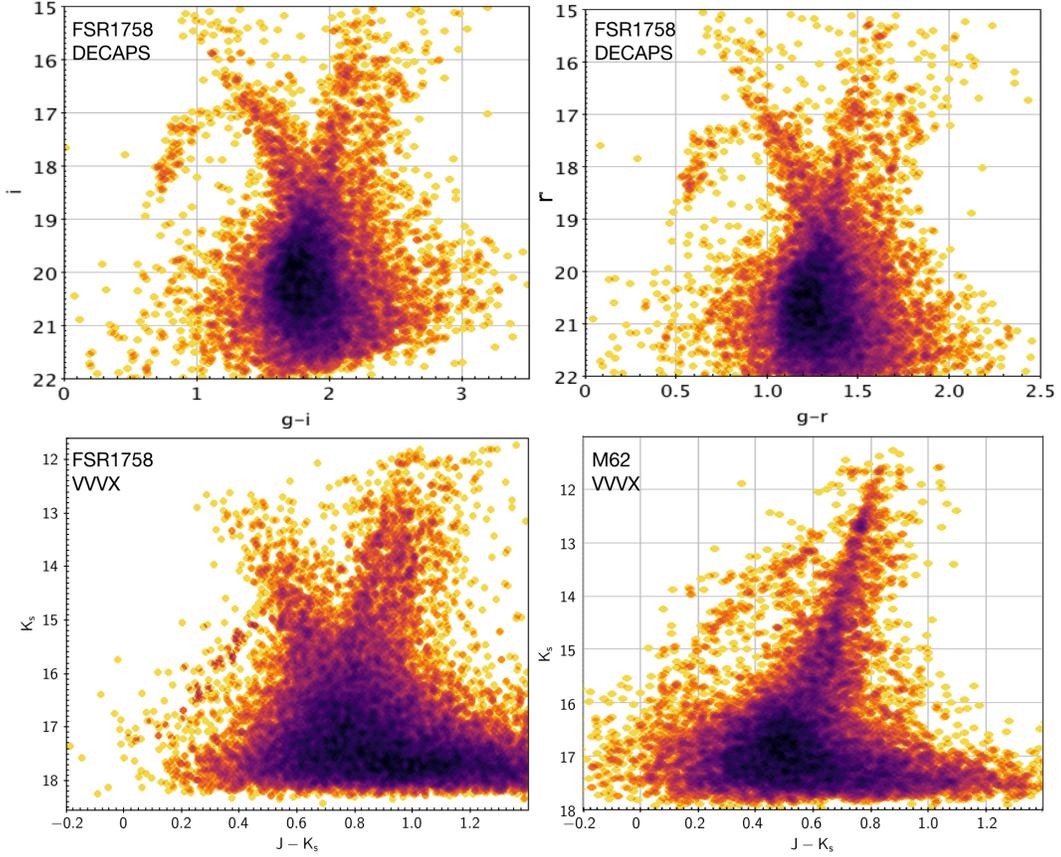}
\caption{ 
Left: DECaPS CMDs for the new GC region: $r$ vs $g-r$ (upper-left), and $i$ vs $g-i$ (upper-right).  
Right: VVVX CMD for the new GC region: $Ks$ vs $J-Ks$
(lower-left), compared with the known GC NGC\,6266 (M62) (lower-right).
}
\label{fig:DECaPS-VVVX-CMDs}
\end{figure}

Figure~\ref{fig:DECaPS-VVVX-CMDs} (top panels) show the optical CMDs using the DECaPS photometry from Schlafly et al. (2018). 
The photometry is very deep, and the cluster red giant branch (RGB) is clearly seen, although heavily contaminated by foreground RGB stars. 
However, the most striking indication of the cluster is the presence of an extended blue horizontal branch (BHB), that is absent in the surrounding fields. 

Figure~\ref{fig:DECaPS-VVVX-CMDs} (bottom left panel) shows the  near-IR CMD within 5' of the cluster centre obtained from PSF photometry of the VVVX tile e682, following the procedure from Alonso-Garc\'ia et al. (2018). 
The cluster RGB and extended BHB are clearly seen, on top of the
foreground disk and background Bulge stars.  
In particular, the cluster RGB is bluer than the Bulge RGB.
A clump of stars located at $Ks=13.4$, $J-Ks=0.87$ can be identified; this feature is likely the red giant branch bump (RGBb). 
This interesting GC CMD feature is related to the evolution of the RGB stars during the first dredge-up, and it is sensitive to metallicity, helium content and mixing efficiency (c.f. Fu et al. 2018).
The locus in a CMD of the RGBb depends on both age and metallicity. Once estimated the cluster metallicity, the location of the RGBb in a CMD helps to determine the age, and viceversa \citep{alves99}.

The optical and near-IR CMDs exhibit a well-populated steep RGB, with no clear red clump, an indication of a metal poor GC, in agreement with the presence of a prominent BHB. 

Schlegel et al. (1998) and Schlafly et al. (2011) determined the extinction in the area to be in the ranges $A_V=3.35:2.89$ (Landolt filters), $A_I=2.11:1.79$ (SDSS filters), and $A_K=0.37:0.32$ (UKIRT filters). 
By comparing the near-IR CMD with known metal-poor clusters, we
obtained  $E(J-Ks)=0.20 \pm 0.03$ mag, in agreement with the
extinction values from  Schlafly et al. (2011) 

Knowing the reddening, we can accurately measure the distance differentially with respect to the known Bulge GCs NGC\,6642 and NGC\,6266, using the VVVX near-IR CMDs.
With respect to NGC\,6642, we measured $\Delta (J-Ks)=0.10$ mag, and $\Delta Ks=0.90$ mag. 
This yields a fainter distance modulus (by about $0.85\pm0.1$ mag) than the GC NGC\,6642, whose distance and metallicity are $D=8.1$ kpc and $[{\rm Fe/H}]=-1.26$ dex, respectively. 
Our first estimate of the distance to $FSR\,1758$ is then $D _{1}=12.0 \pm 0.5$ kpc. Similarly, with respect to NGC\,6266,  we measured $\Delta (J-Ks)=0.10$ mag, and $\Delta Ks=0.95$ mag. 
This gives a fainter distance modulus than the GC NGC\,6266, which has $D=6.8$ kpc and $[{\rm Fe/H}]=-1.18$ dex. 
Then, our second estimate is $D_{2}=11.0 \pm 0.5 $ kpc.
We can take the average of the former two values as the distance to FSR\,1758, i.e.  $D_{FSR 1758}=11.5 \pm 1.0$ kpc. 
The error is estimated as the sum of the individual errors in order to include possible systematic differences.
The alignment of the RGB and the BHB with these comparison GCs is remarkably good, which gives added confidence to this determination. 
We note that the GDR2 parallaxes in the distant and crowded cluster field would be unreliable. 

We repeated the same procedure using the Gaia CMD, and considering the clusters studied by Babusiaux et al. (2018) as references.
Using this optical CMD we find a magnitude difference of $17.1 \pm 0.1$ mag, and a reddening $E(BP − RP ) = 0.90 ± 0.05$ mag. 
Using the extinction ratio from Andrae et al. (2018) yields $AG = 2.0\,E(BP − RP ) = 1.8$ mag. 
This gives an absolute distance modulus of $(m − M)_0 = 15.30 ± 0.10$ mag, equivalent to a distance $D = 11.5 ± 0.5$ kpc.
The Gaia CMD distance is then very consistent with the determination using the near-IR CMD. 
We note here that the fitting uncertainty of about 0.1 mag is quite optimistic, because it does not consider the uncertainty in the reddening law. 
Therefore, as mentioned before, a distance error of the order of 1 kpc must be considered.
As a conclusion, FSR\,1758 is located on the far side beyond the Bulge.
Its corresponding Galactocentric distance is $R_G=3.7$~kpc,  and its height below the plane is $z=-760$~pc.

\begin{figure}[ht]
\centering
\includegraphics[height = 55 mm]{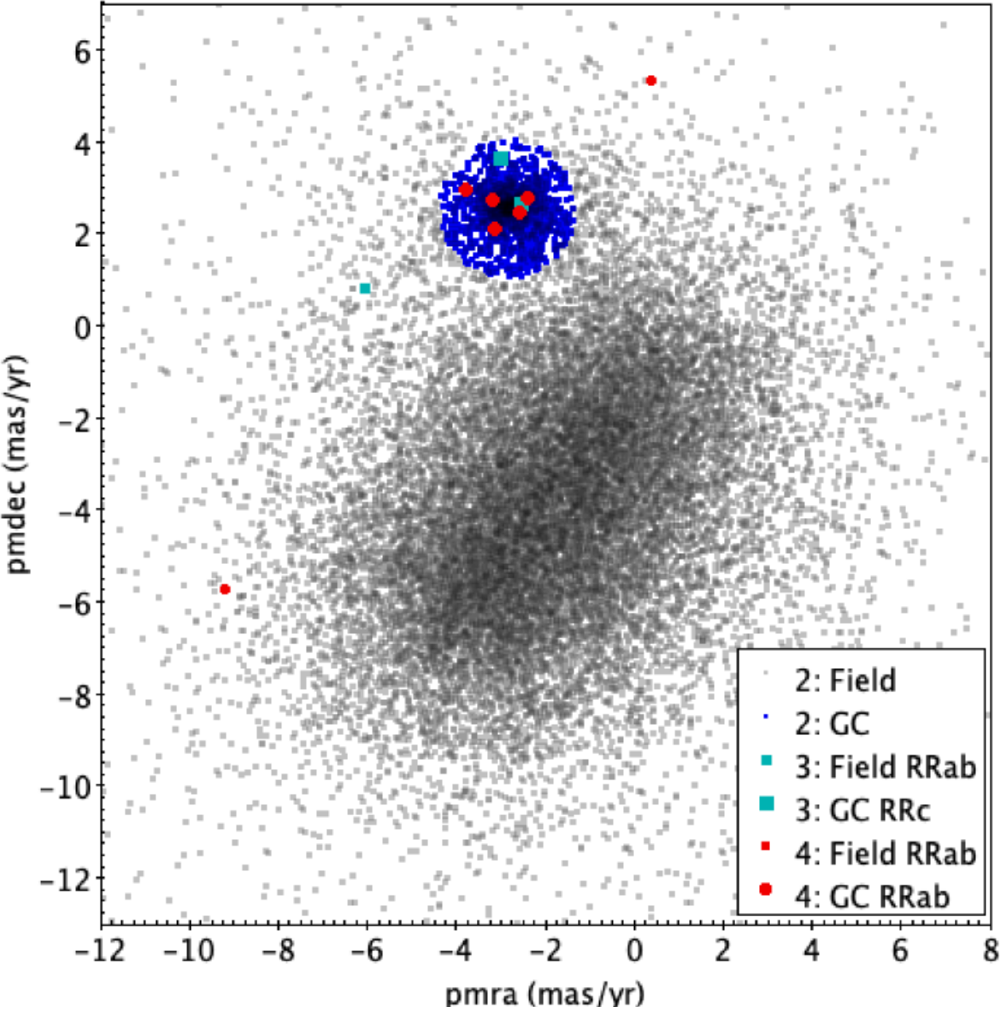}
\includegraphics[height = 55 mm]{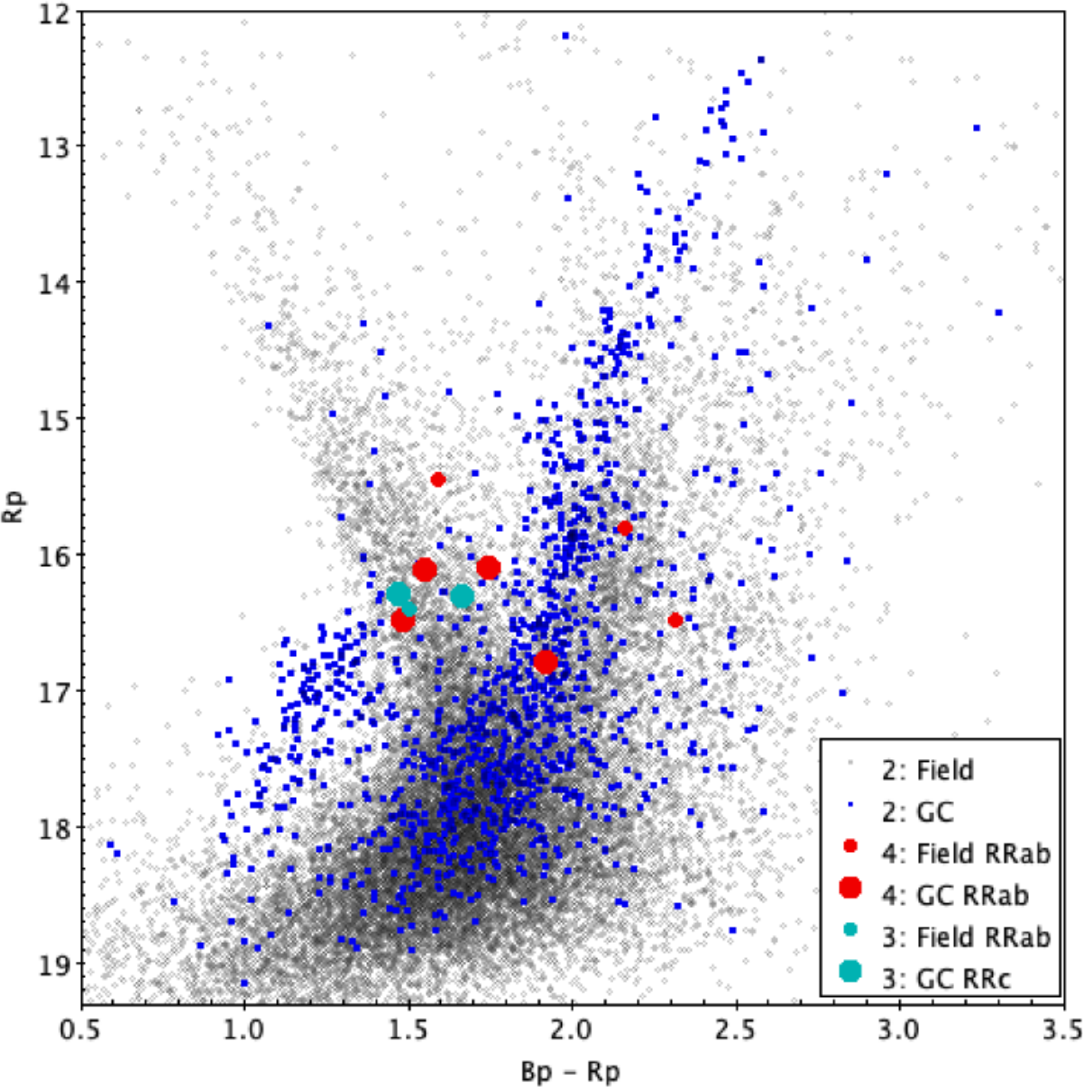}
\includegraphics[height = 55 mm]{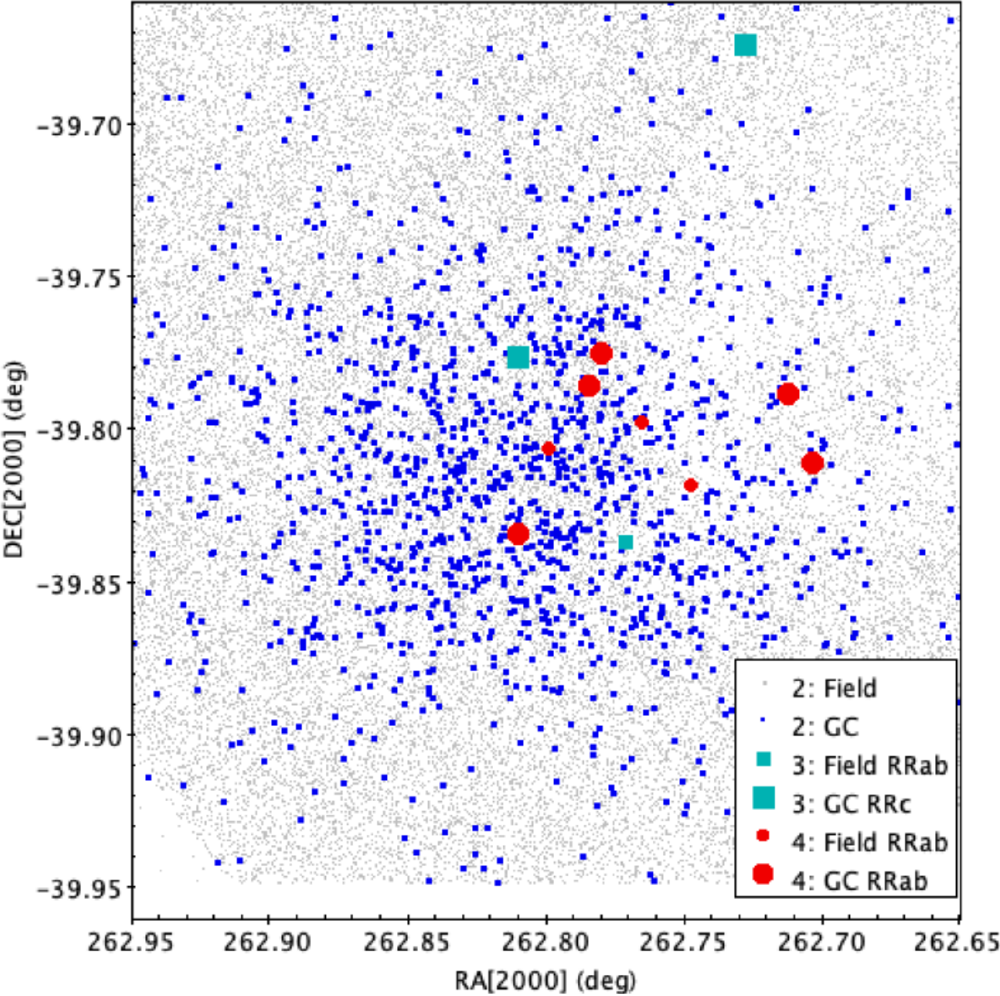}
\caption{ 
Left panel: Gaia PM diagram for the cluster region, with the selected stars around the $(\mu_\alpha,\mu_\delta)=(-2.85,2.55)$ mas marked in blue. 
Center panel: Gaia CMD of all stars in the region (gray dots),
compared with the PM selected GC members (blue dots).
Right panel: Spatial distribution of stars, probable members marked in blue.  
The RRLyr variable stars are also shown as red circles (type RRab),
and cyan (type RRc). Large symbols are probable members of the
cluster, while small symbols are stars beloging to the Bulge.
}
\label{fig:GDR2}
\end{figure}

Figure~\ref{fig:GDR2} (middle panel) shows the optical Gaia CMD, decontaminated using PM data.
Again, the cluster RGB and extended BHB are clearly visible. 
The cluster RGBb is also well defined, located at $G=16.9$, $BP-RP=2.0$.
The RGB tip is located at Gaia magnitude $G=13.5$, bright enough for future spectroscopic followup.   
Figure~\ref{fig:GDR2} also shows the cluster PM (left panel), which is strikingly different from nearby field stars. 
This characteristic led Cantat-Gaudin et al. (2018) to claim the discovery of a new GC.  
The longitudinal component of the PM is close to zero, and indeed this cluster has a PM indicating it is plunging into the disk. 

The wide wavelength coverage available allows to constrain the
cluster metallicity. 
Once the reddening is known, the metallicity can be readily estimated from the optical and near-IR CMDs.  
Compared to other well-known GCs (e.g. Babusiaux et al. 2018), 
the Gaia CMD of FSR\,1758 resembles those of M3 and M13, thus suggesting a metallicity close to $[{\rm Fe/H}]=-1.5$ dex. 
Using the RGB location interpolated in the optical CMDs from Babusiaux et al. (2018), and the near-IR CMDs from Valenti et al. (2004), we estimated a value of $[{\rm Fe/H}]=-1.5 \pm 0.3$ dex for the metallicity of this cluster. 

A concentration of RR~Lyrae variables is also evident in the cluster region. 
From the OGLE catalog by Soczynski et al. (2014), we found eight fundamental RR~Lyrae pulsators (RRab), plus three first overtone RR~Lyrae pulsators (RRc) within 0.15 deg (9') from the cluster center. 
The centroid of the RR~Lyrae distribution is located to the West of
the centroid of the BHB distribution, probably due to a differential reddening effect. 
Table~\ref{tab:table1} lists the RR~Lyr stars including their OGLE IDs, types, periods, GDR2 equatorial coordinates and PMs (in J2000), OGLE mean V and I-band and GDR2 magnitudes, extinctions $A_I$, distance from the cluster center, and probable cluster membership.
We select the most likely cluster members based on the sky positions, PMs, and mean magnitudes and positions in the CMDs.
The mean period for the five RRab members is $<P>=0.684$ d. 
As a consequence, we classify this cluster as an Oosterhoff type II, which is consistent with being metal-poor like, for example, $\omega$~Cen (Clement \& Rowe 2000).

RR~Lyrae can be used to determine GCs distances even in deeply reddened cases (e.g. Alonso-Garc\'ia et al. 2015,  Minniti et al. 2018b). 
We measured the distance to FSR\,1758 differentially with respect to $\omega$~Cen following Braga et al. (2018), who relied on the period-luminosity-metallicity relations by Marconi et al. (2015). 
Considering all eight RRab listed in Table~\ref{tab:table1}, we obtained a mean distance modulus of $<m-M_0>=15.16 \pm 0.3$ mag, equivalent to a distance $D = 10.8 \pm 1.0$ kpc.
On the other hand, using only the five RRab that are most probably cluster members we obtained $<m-M_0> = 15.02 \pm 0.3$ mag, equivalent to $D = 10.1 \pm 1.0$ kpc. 
Although the former estimate has a larger scatter due to the large individual reddening corrections, it is still in agreement with the distance determination based on the optical and near-IR CMDs.

\section{The Nature of FSR\,1758: A Giant GC or Dwarf Galaxy?}
\label{sec:sec5}

Figure~4 shows the radial profile and spatial extension of FSR\,1758, obtained combining DECaPS and GDR2 data.
We isolated stars with PMs similar to the cluster, within $1.2$ mas from $\mu_{\alpha} = -2.85$ mas yr$^{-1}$, and $\mu_{\delta} = 2.55$ mas yr$^{-1}$, and parallaxes smaller than 0.3 mas, in order to avoid most of the foreground stars.    
The DECaPS star counts in $i$-band (shown in the left panel) indicates that the cluster stellar density joins the field at large radii from the center ($R > 15'$). 
The differential extinction in the field produces a structured radial profile. 
The fitting of a King profile (King 1962, 1966) gives a core radius $R_c=0.050 \pm 0.004$ deg (about $10 \pm 1$ pc) and a tidal radius $R_t=0.78 \pm 0.22$ deg (about $150 \pm 45$ pc). 
The derived value of $R_t=150$ pc should be taken with caution due to the presence of strong differential reddening in the area.
Although the structure  of FSR\,1758 determined from the star counts yields a concentrated radial profile typical of a GC, with a concentration index $c = 1.20 \pm ^{0.19}_{0.13}$, it suggest an extended nature of the cluster.
Figure 4 shows that this object is indeed very extended, possibly even larger than $\omega$~Cen, the most massive GC in the Milky Way (e.g. Meylan 1987, Ferraro et al. 2006, Harris 1996, 2010), with a tidal radius of $R_t=45'$ (Trager, King \& Djorgovski 1995), and a mass of $4\times 10^6 ~M_{\odot}$ (D'Souza \& Rix 2013).  
It is also a significantly flattened GC (White \& Shawl 2000, Chen \& Chen 2010). 
The appearance of FSR\,1758 is also very flattened, like $\omega$ Cen, although this issue needs further investigation since the dark cloud located to the North-East of the cluster centre could be significantly affecting this result.  
The total cluster luminosity is very difficult to estimate in the presence of the high background and heavy differential reddening.
A lower crude estimate was obtained using the DECaPS photometry in the $i$-band. 
Coadding up all stars within 0.1 deg from the GC center (to avoid further field contamination), after accounting for the background taken in four different fields of similar area surrounding the cluster, assuming a distance of $11.5 \pm 1.0$ kpc, and an extinction of $A_i= 1.79$ mag from Schlafly et al. (2011), we obtain a total $i$-band absolute magnitude brighter than $M_i< -8.6 \pm 1.0$ mag. 
This is a very bright GC indeed, and we emphasize that this total magnitude is only a lower limit because of incompleteness and differential reddening.
A detailed study of reddening is needed in order to improve the values of astrophysical parameters allowing a more robust comparison  with $\omega$ Cen.

\begin{figure}[ht]
\centering
\includegraphics[height = 70 mm]{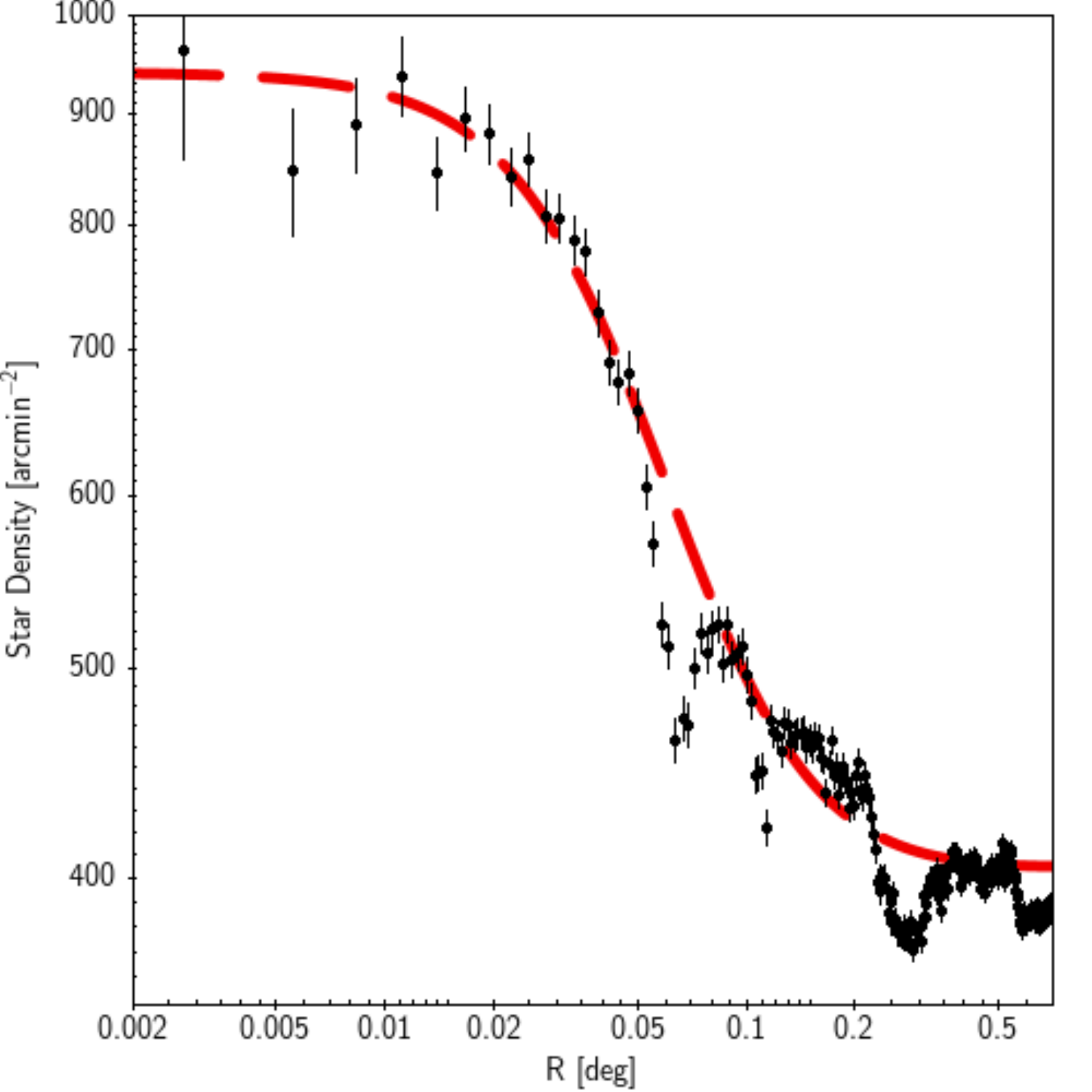}
\includegraphics[height = 70 mm]{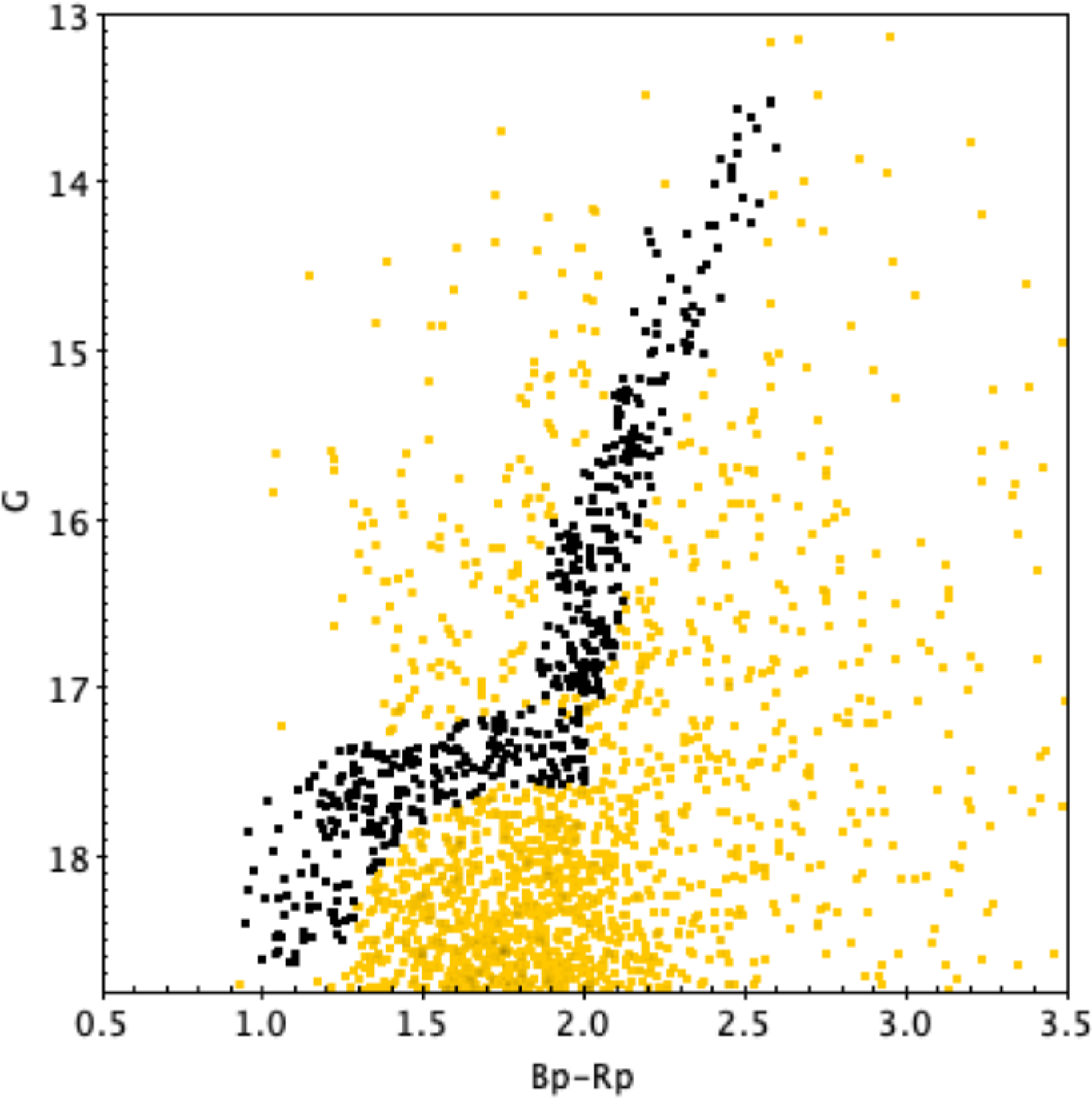}
\includegraphics[height = 70 mm]{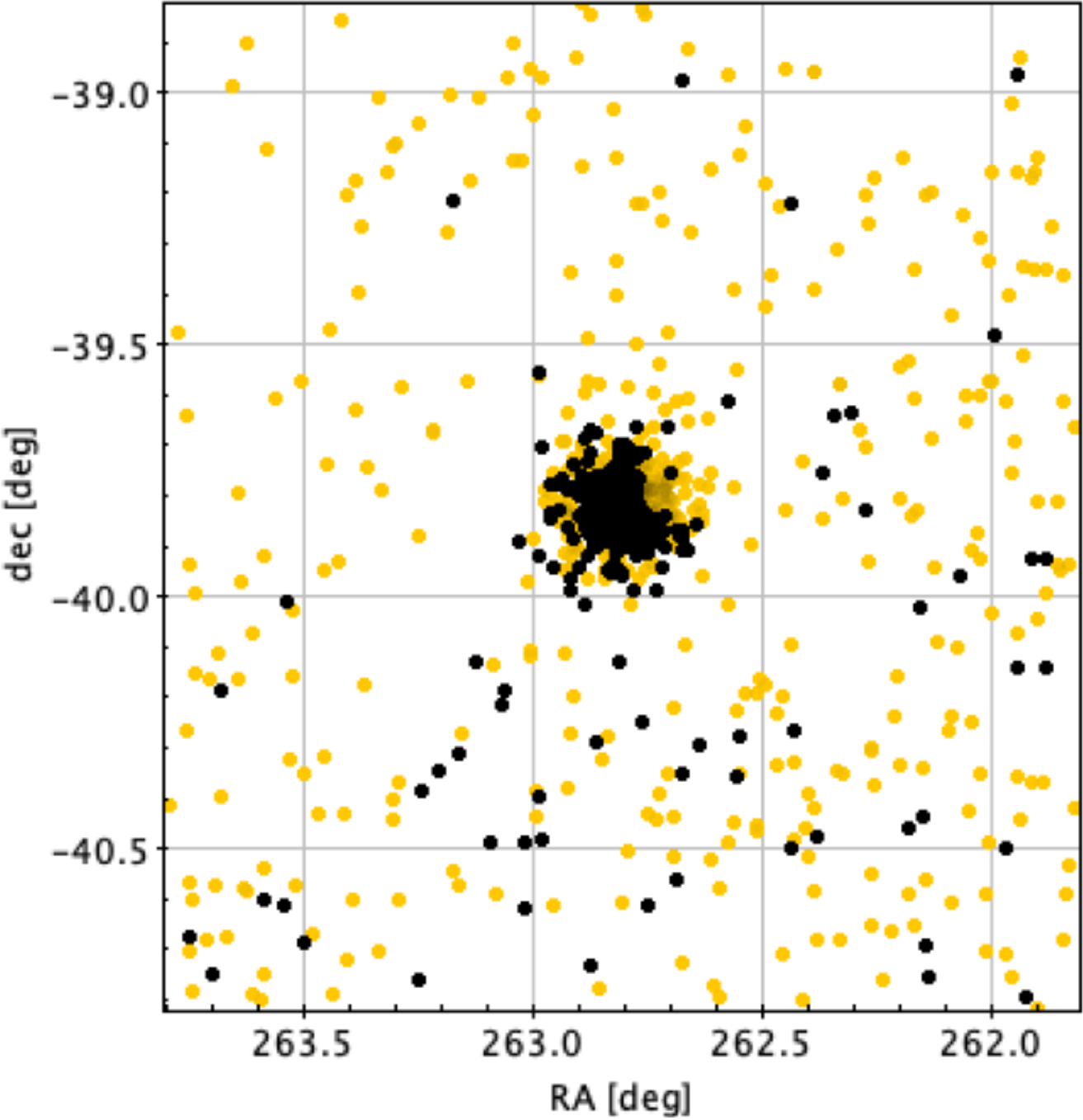}
\includegraphics[height = 70 mm]{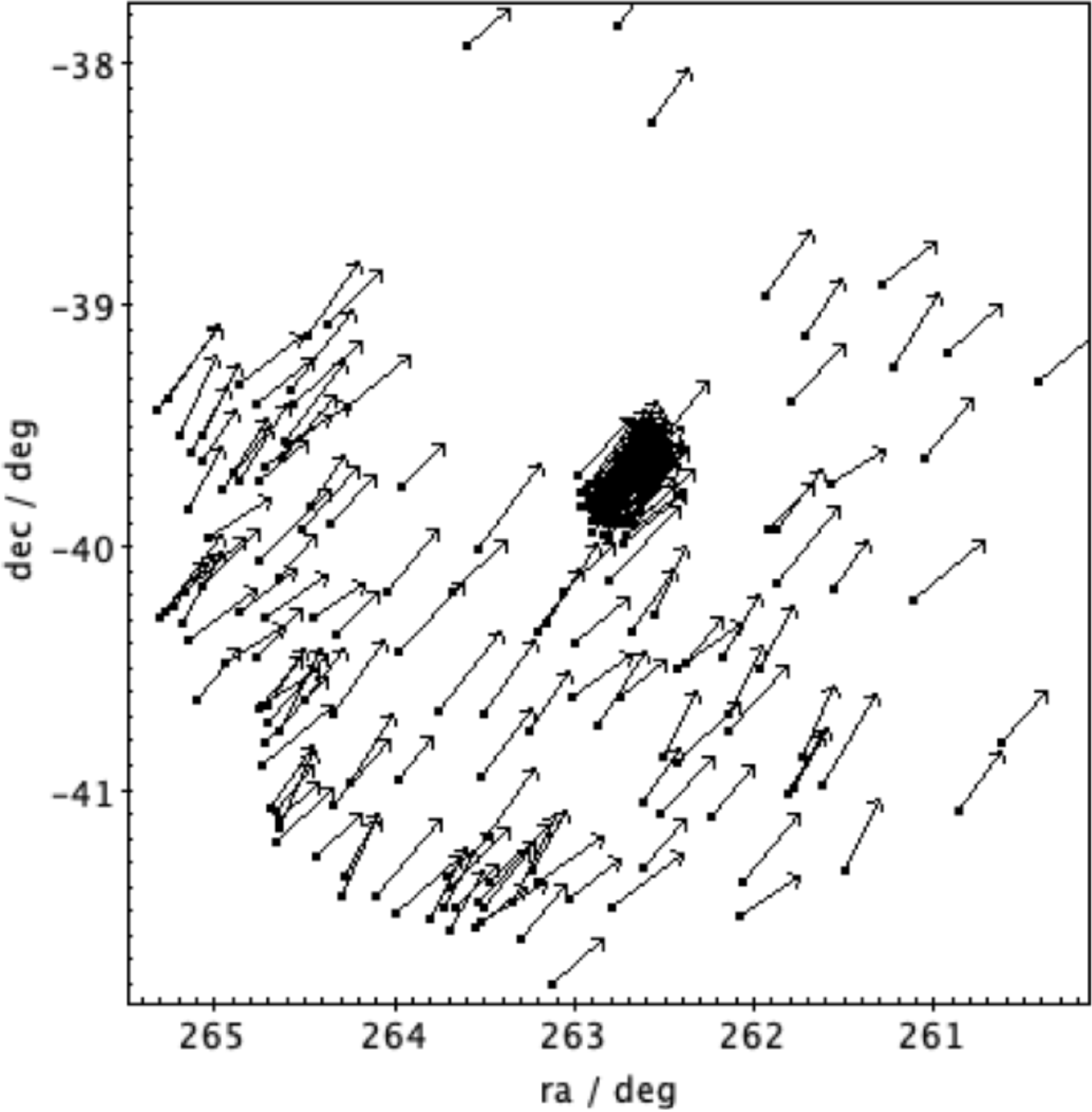}
\caption{
Top-left panel: Stellar radial density profile of stars using DECaPS $i-$-band photometric data (black dots) and King's profile fitting (red line). 
This shows that the cluster size is extended beyond $0.5$ deg.
Top-right panel: GDR2 CMD of selected (by PM and parallax) stars in the area of $2^\circ$ radius, with BHB and RGB stars highlighted in black.
Bottom-left panel: Spatial distribution of Gaia selected BHB and RGB stars over a $1^\circ \times 1^\circ$ field. 
Note that some faint stars probably belonging to the cluster are missed due to the selection criteria applied to separate BHB and RGB stars from others.
Bottom-right panel: Spatial distribution of common PM stars, showing a possible larger structure extended a few degrees to the South-east. 
The Galactic Plane is located toward the North-west. 
}
\label{fig:hvs_stream}
\end{figure}

The large size of the cluster is confirmed using stellar tracers like BHB stars.
Note that BHB stars are particularly sensitive to extinction, and therefore the empty patches just represent high extinction fields.  
Because there was no clear edge to the distribution of BHB stars, we decided to explore the spatial distribution of comoving stars using the GDR2 PMs.
We then searched for associated streaming motions within two degrees of the cluster, using the same set of stars isolated from GDR2.
Then, we selected the stars that lie in the main locus of the cluster RGB and BHB, with an additional constraint $G < 18.8$. 
The result, shown in the right panel of Figure~\ref{fig:hvs_stream}, reveals an asymmetric source distribution, with the potential comoving stars located preferentially to the South-East. 
This suggests that FSR\,1758 may be the nucleus of a dwarf galaxy, which we tentatively name as {\bf Scorpius} dwarf galaxy, joining in this category to the GCs $\omega$ Cen (Meylan et al. 2001), M54 (Monaco et al. 2005), and M31-G1 (Gregg et al. 2015). 
In particular, some of the BHB stars in the outskirts of the field studied may belong to the body of this putative dwarf galaxy, in analogy with the BHB stars found by Monaco et al. (2003) in the field surrounding the GC M54, which is now known to be the nucleus of the Sgr dwarf galaxy. 
It could also be one of the putative primordial bulge building blocks such as Terzan 5 (Ferraro et al. 2009). 

Also interesting in this regard is the fact that FSR\,1758 appears not to fit to the well-defined size-metallicity and size-Galactocentric radius relations for the Galactic GCs (e.g. Vanderbeke et al. 2015), but to lie far above the mean correlations observed between these parameters. 
A thorough comparison of its properties with those of dwarf galaxies is beyond the scope of this paper. For example, in absence of radial-velocity data, it is not possible to determine its mass-to-light ratio. 
Given its metallicity, if it was a dwarf galaxy, its stellar mass would be of the order of $10^7~\rm{M_{\odot}}$, according to the mass-metallicity relation for dwarf galaxies \citep{kirby13}. 
This relation could remain valid even if the object is tidally stripped.

Since FSR\,1758 is close in projection to other two GCs belonging to the Bulge, Ton~2 and NGC~6380, we explored the possibility of a real association with any of them.
We found that they all share similar PMs; all three clusters might be plunging
onto the Galactic disk.

With an angular separation of only 0.8 deg and a probable physical separation of about $600$ pc (perhaps even smaller considering the large errors in the distances), the association of FSR\,1758  with  NGC\,6380 (Harris 1996, 2010) is strong.
They might be a binary cluster or part of a larger structure like a dwarf galaxy. 
However, their compositions are different by ~0.7 dex, with NGC~6380 being
more metal-rich. 
The angular separation between Ton\,2 and FSR\,1758 is 1.4 deg. The distance to the former cluster is believed to be  $D=8.2$ kpc, although it is not well established, and its metallicity was determined to be $[{\rm Fe/H}]=-0.70$ dex.
Given the different metallicities, a putative association is not strongly supported. 
In addition, the radial velocities of Ton\,2 and NGC\,6380 are very different, $-182$ km\,s$^{-1}$ and $-4$ km\,s$^{-1}$,respectively, so these clusters are definitely not associated with each other.

Spectroscopic data of a number of stellar members of FSR\,1758 are essential, not only to compute the radial velocity, necessary to determine the orbital path of the object, but also to derive information about its origin and potential association with any of the known GCs, and to obtain metallicity estimates. 
Combining the velocity dispersion and metallicities for various stars will allow to definitively determine the nature of this object by estimating its dark matter content and whether or not it possesses a metallicity spread.


\section{Conclusions}
\label{sec:sec6}

FSR\,1758 is a new globular cluster (GC) recently discovered serendipitously in the Milky Way Bulge. 
In this paper we used combined data from  the Gaia, DECaPS and VVVX surveys to determine the cluster physical parameters for the first time, including its position, distance, reddening, size, metallicity, and mean proper motion.  
The cluster is centered at equatorial coordinates $RA=17:31:12$, $DEC=-39:48:30$ (J2000), and Galactic coordinates $l=349.217$ deg, $b=-3.292$ deg. 
A distance of $D=11.5\pm 0.5$ kpc was estimated using the optical and near-IR CMDs, and confirmed with OGLE RR Lyrae variable stars.  
From the CMDs, extinction values were also measured, resulting in $E(J-Ks)=0.20 \pm 0.03$ mag, and $E(BP-RP)=0.90\pm 0.05$ mag. 
The metallicity of FSR\,1758 was estimated to be $[{\rm Fe/H}]=-1.5 \pm 0.3$ dex, based on the appearance of the RGB in the optical and near-IR CMDs. 
Finally, its derived mean proper motion,  $\mu_{\alpha} = -2.85$ mas yr$^{-1}$ and $\mu_{\delta} = 2.55$ mas $^{-1}$, indicates that this cluster is plunging onto the Galactic plane.
The acid test for this cluster will be to obtain spectra for a number of members.
The measurement of radial velocities which, combined with the distance and proper motion information, will give us the orbital parameters for the individual stars, and the chemical information will allow us to search for any metallicity variation or the presence of multiple populations.

The stellar density distribution shows that FSR\,1758 is a very large cluster, with a core radius $R_c=10$ pc, and a large tidal radius of about $R_t = 150$ pc, with a lower limit for the $i$-band luminosity of about $M_i < -8.6$.
The inspection of a larger field of view of several degrees around the cluster led to the detection of extended streaming motions, i.e. a collection of stars with characteristics similar to the BHB and RGB stars of the cluster, with coherent PMs, and moving parallel to the cluster. 
This points to the hypothesis that this object is actually much larger, possibly the nucleus of a dwarf galaxy.

The census of Galactic GCs is not complete, and several low-luminosity GCs may still be missing in the Galactic Bulge (Bica et al. 2016, Minniti et al. 2017, Ryu \& Lee 2018, Camargo 2018). 
The discovery of FSR\,1758 represents a paradigm shift, as it clearly shows that not only low-luminosity GCs may be missing in the Bulge, but also some quite luminous and massive ones.  
Future searches to be carried out with the Large Synoptic Survey
Telescope (LSST) \citep{ivezic08}, or with the Wide Field Infrared
Survey Telescope (WFIRST) \citep{spergel15,stauffer18} might detect more GCs hidden in this region. 
Finally, the recently started multi-epoch observation campaign for the VVVX survey would significantly extend the areal coverage, allowing to search for  more RR~Lyrae variable stars likely associated  with this new large GC.

\acknowledgments 

We thank the reviewer, who did a thorough reading, which helped to substantially improve the manuscript.
We gratefully acknowledge  data from the ESO Public  Survey program
ID179.B-2002  taken with  the  VISTA telescope,  and  products from
the Cambridge Astronomical Survey Unit (CASU).  
We have made use of tools from Aladin/Simbad at the Centre des Donnees
Stelaires (CDS) Strassbourg, and TopCat (Taylor 2005).
R.H.B. thanks support from DIULS project PR18143, and the ''Big Data
and Cosmography: the skyscape of large astrophysical surveys'' project.
D.M. and D.G.  gratefully acknowledge support provided by the BASAL  Center
for  Astrophysics and  Associated  Technologies  (CATA) through grant
AFB-170002.
D.G. also acknowledges financial support from the Direcci\'on de
Investigaci\'on y Desarrollo de la Universidad de La Serena through
the Programa de Incentivo a la Investigaci\'on de Acad\'emicos
(PIA-DIDULS). D.M. acknowledges the  Ministry for the  Economy,
Development and Tourism,Programa Iniciativa  Cient\'ifica Milenio
grant IC120009, awarded to the Millennium Institute  of Astrophysics
(MAS), and from FONDECYT Regular grant 1170121. 
J.A-G. acknowledges support by FONDECYT Iniciation grant 11150916, and
by the Ministry of Education through grant ANT-1656.  
A.M. acknowledges support from FONDECYT Regular grant
1181797. F. G. acknowledges support from FONDECYT Regular grant
1181264. F.G and A.M aknowledge funding from the Max Planck Society
through a ''Partner Group'' grant.


\clearpage

\begin{deluxetable}{llccccccccccccccl}
\tabletypesize{\scriptsize}
\setlength{\tabcolsep}{0.02in}
\tablewidth{0pt}
\tablecaption{RR Lyrae pulsators in the field of FSR\,1758\label{tab:table1}}

\tablehead{
\colhead{ID}  &
\colhead{Type}&
\colhead{P(days)} & 
\colhead{RA}  & 
\colhead{DEC} &
\colhead{$\mu_\alpha$} &
\colhead{$\mu_\delta$} &
\colhead{$V$}    &
\colhead{$I$}    & 
\colhead{$G$}    &
\colhead{$B_P$}  &
\colhead{$R_P$}  &
\colhead{$A_I$}  &
\colhead{$R$(")} &
\colhead{PM}     &
\colhead{CMD}    &
\colhead{Notes}
\\
}
\startdata
OGLE-BLG-RRLYR-00882 & RRab & 0.6396 & 262.70335 & -39.81101 & -3.195 &  2.721 & 18.771 & 16.735 & 17.877 & 18.710 & 16.787 & 1.68 & 267.5 & Y & Y & GC      \\       
OGLE-BLG-RRLYR-00883 & RRab & 0.6606 & 262.71236 & -39.78844 & -2.411 &  2.777 & 18.540 & 16.735 & 17.918 &        &        & 1.73 & 252.7 & Y & Y & GC      \\             
OGLE-BLG-RRLYR-00887 & RRab & 0.5183 & 262.74721 & -39.81809 &  2.174 &-21.062 & 18.093 & 15.816 & 16.988 & 17.966 & 15.803 & 1.63 & 150.0 & N & N & Bulge   \\       
OGLE-BLG-RRLYR-00889 & RRab & 0.5762 & 262.76517 & -39.79741 &  0.397 &  5.344 & 18.999 & 16.932 & 18.153 & 18.792 & 16.474 & 1.66 & 104.0 & N & N & Unknown \\       
OGLE-BLG-RRLYR-00891 & RRab & 0.5504 & 262.77940 & -39.77487 & -3.144 &  2.113 & 17.788 & 16.241 & 17.362 & 17.960 & 16.479 & 1.68 & 133.3 & Y & Y & GC      \\        
OGLE-BLG-RRLYR-00893 & RRab & 0.7642 & 262.78404 & -39.78579 & -3.779 &  2.969 & 17.649 & 15.983 & 17.035 & 17.658 & 16.107 & 1.65 &  92.4 & Y & Y & GC      \\       
OGLE-BLG-RRLYR-00894 & RRab & 0.7561 & 262.79889 & -39.80606 & -9.222 & -5.744 & 17.237 & 16.013 & 16.879 & 17.047 & 15.456 & 1.65 &   8.7  & N & N & Bulge   \\       
OGLE-BLG-RRLYR-00896 & RRab & 0.8062 & 262.80977 & -39.83407 & -2.597 &  2.440 & 17.572 & 16.046 & 17.157 & 17.834 & 16.091 & 1.57 & 96.5 & Y & Y & GC      \\
OGLE-BLG-RRLYR-00885 & RRc  & 0.3480 & 262.72721 & -39.67458 & -2.981 &  3.610 & 18.049 & 16.389 & 17.531 & 17.960 & 16.297 & 2.05 & 522.0 & Y & Y & GC      \\       
OGLE-BLG-RRLYR-00890 & RRc  & 0.3214 & 262.77060 & -39.83700 & -6.044 &  0.811 & 18.046 & 16.434 & 17.403 & 17.908 & 16.407 & 1.77 & 131.3 & N & Y & Unknown \\       
OGLE-BLG-RRLYR-00895 & RRc  & 0.3292 & 262.80971 & -39.77662 & -2.531 &  2.640 & 17.650 & 16.387 & 17.429 & 17.763 & 16.296 & 1.83 & 117.3 & Y & Y & GC      \\    \enddata

\bigskip
\tablecomments{$V$, $I$ values are from Soszynski et al. (2011).\\
Positions, proper motions, $G$, $B_p$, and $R_p$ values are from GDR2.\\
Typical OGLE photometric errors are $\sigma_{V} = 0.01$ mag, and $\sigma_{I} = 0.01$ mag.\\ 
Typical period errors are $\sigma_P=0.00001$ days.\\ 
Typical proper motion errors: $\mu_\alpha=0.43$ mas, and $\mu_\delta=0.32$ mas.\\
Extinction values $A_I$ are from Schlafly et al. (2011).}
\end{deluxetable}

\end{document}